\documentclass{article}

\usepackage{PRIMEarxiv}

\usepackage[utf8]{inputenc} 
\usepackage[T1]{fontenc}    
\usepackage{hyperref}       
\usepackage{url}            
\usepackage{booktabs}       
\usepackage{amsfonts}       
\usepackage{nicefrac}       
\usepackage{microtype}      
\usepackage{lipsum}
\usepackage{fancyhdr}       
\usepackage{graphicx}       
\graphicspath{{media/}}     
\usepackage{xcolor}
\usepackage{float}
\usepackage{multirow}
\usepackage{array}
\usepackage{setspace} 
\usepackage{caption}   
\usepackage{subcaption}
\usepackage{amssymb}
\usepackage{titling} 
\usepackage[ruled,vlined]{algorithm2e}
\usepackage{amsmath}

\title{\LARGE Topology-aware Preemptive Scheduling for Co-located LLM Workloads
}

\author{
  Ping Zhang, Lei Su, Jinjie Yang, Xin Chen \\
  Baichuan-Inc \\
  \texttt{\{zhangping, sulei, yangjinjie, chenxin\}@baichuan-inc.com} \\
}

\date{}

\begin{document}
\maketitle

\begin{abstract}
Hosting diverse large language model workloads in a unified resource pool through co-location is cost-effective. For example, long-running chat services generally follow diurnal traffic patterns, which inspire co-location of batch jobs to fulfill resource valleys between successive peaks, and thus to saturate resource allocation in cluster-wide scope. These heterogeneous workloads often have different business priorities, and therefore preemption can be leveraged for resource elasticity. However, workloads often have distinct topology preferences as well. The resources released by lower-priority instances may fail to meet the requirements of high-priority online services which are usually latency-sensitive. The root cause behind such mis-match is a lack of topology awareness of resource scheduler, especially during preemption. To bridge this gap, we develop a fine-grained topology-aware method for preemptive scheduling of hybrid workloads. The method ensures that the resources freed by preempted tasks adhere to the topological affinity needs of high-priority preemptors in a guaranteed or best-effort manner. This dynamic alignment significantly increases the efficiency of preemption and improves overall scheduled performance for LLM workloads by $55\%$.
\end{abstract}


\section{Introduction}
\label{sec:intro}
The rapid deployment of services powered by large language models (LLMs) has gained significant attention due to their versatility across a wide range of applications \cite{hadi2023survey}. With their surge in popularity, service providers that host self-managed LLMs are facing increasing pressure to optimize resource utilization, especially of GPU servers, which constitute a substantial portion of operational costs.

Unlike the pre-training phase, which typically runs a single large-scale job on dedicated or exclusive clusters, where resource allocation is straightforward \cite{duan2024efficient}, the inference serving stage of LLMs benefits from co-locating multiple workloads in a unified resource pool. For example, the online chat services such as ChatGPT\footnote{https://openai.com/chatgpt/} and Baixiaoying \footnote{https://ying.baichuan-ai.com/chat} usually exhibit diurnal patterns in traffic \cite{wang2024towards}, inspiring auto-scaling to ensure performance demands in peaks while to save cost in valleys. By co-locating offline workloads such as data processing jobs and/or offline inference tasks, the resource valleys between successive peaks of online services can be then padded, improving resource efficiency and reducing cost \cite{xiang2023godel}.

While co-location improves resource utilization, it introduces scheduling complexities due to the heterogeneous nature of diverse LLM workloads, and complicates the maintenance of consistent performance, especially during preemption, in which lower-priority tasks are evicted in order to guarantee the resource demands of higher-priority services when resources are insufficient. Concretely, the scheduling cycle in a co-located cluster is often driven by auto-scaling of online services, responding to fluctuations in user request volume. Those online services are usually latency-sensitive, requiring a high performance placement on the GPU server according to its hardware topology affinity \cite{amaral2017topology}.
However, the resources freed by preempted instances often do not satisfy the topological affinity requirements of online services. This mismatch stems from the lack of topology awareness in preemptive scheduling, frequently resulting in deployment failures or suboptimal performance for the online services (i.e., the preemptors). Consequently, this leads to inefficient resource elasticity and underutilization.

To address this gap, we propose a fine-grained, topology-aware preemption method specifically designed for scheduling hybrid LLM workloads. Our approach ensures that when lower-priority jobs are preempted, the resources they release are allocated in a manner that adheres to the topological affinity requirements of high-priority services, such as maintaining socket affinity, minimizing NUMA (non-uniform memory access \cite{lameter2013numa}) node crossings and preserving CPU-GPU locality. This best-effort alignment optimizes the placement of critical workloads, thereby enhancing the efficiency of preemptive scheduling. As a result, our method significantly improves overall resource utilization in GPU clusters.

This paper is structured as follows. Section ~\ref{sec:background-motivation} details the research background and clarifies our motivation. Section ~\ref{sec:topology-aware-preemption} introduces the proposed topology aware preemption and Section ~\ref{sec:implementation} shows critical implementations. In Section ~\ref{sec:evaluation}, both large-scale simulated experimental results and evaluation on real production clusters are reported and discussed. Section ~\ref{sec:related-work} reviews existing work related to topology-aware scheduling and Section ~\ref{sec:conclusion} concludes the paper.

\section{Background and Motivation}
\label{sec:background-motivation}

\subsection{Distributed LLM Serving}
\label{subsec:distributed-llm-serving}
Existing work has focused on the high cost optimization of LLM serving by improving the inference engine \cite{griggs2024m}. The popular open-sourced serving engines include FasterTransformer \cite{nvidia2021fastertransformer}, vLLM \cite{kwon2023efficient}, FlexGen \cite{flexgen2023}, TGI \cite{huggingface2023textgen}, DeepSpeedInference \cite{deepspeed2022inference}, TensorRTLLM \cite{nvidia2023tensorrtllm} et al. Readers are refereed to \cite{miao2023towards} for a comprehensive survey. This line of work targets on single-instance optimization, and thus is the basis of efficient multi-instance serving.

To manage multi-instance distributed services across a machine cluster, containerization and Kubernetes orchestration have become the de-facto standard \cite{xiang2023godel}. Kubernetes offers robust scalability and flexible orchestration, enabling efficient management of containerized LLM services by automating deployment and scaling for diverse workloads. Influential providers such as OpenAI \footnote{https://openai.com/index/scaling-kubernetes-to-7500-nodes/} are hosting LLM chat services using Kubernetes.

\begin{figure}
  \centering
  \includegraphics[width=0.55\textwidth]{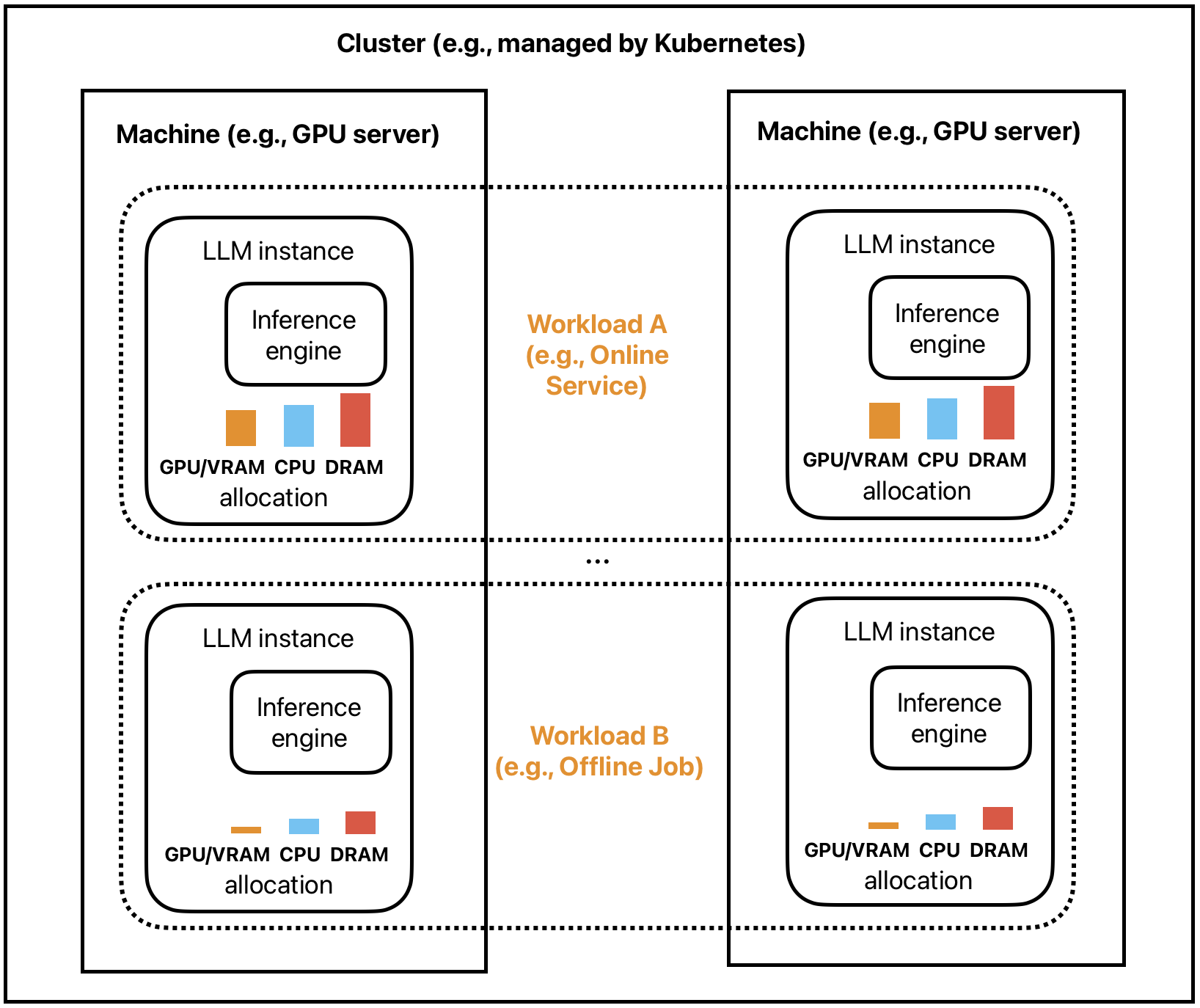}
  \caption{Distributed LLM serving with workload co-location from the cluster view. Hereby, we take an online LLM inference service and an offline LLM inference job as illustrative examples of co-location. Other types of workloads are also considered for similar optimizations.}
  \label{fig:high-level-view-co-location}
\end{figure}

This paper focuses on multi-instance LLM serving with diverse workload co-location. Figure ~\ref{fig:high-level-view-co-location} illustrates the co-location from a high-level deployment view. Each workload consists of multiple LLM instances spreading on GPU servers across the cluster. The instances within the same workload have identical resource allocations, including GPU, CPU, and DRAM (analogous to Pods within the same Deployment in Kubernetes). In contrast, instances from different workloads have distinct resource configurations tailored to their specific requirements. From the view of a single machine, multiple instances from different workloads are co-located locally and isolated by, i.e., containers.

\subsection{Hardware Topology Preference}
\label{subsec:hardware-topology-preference}
\paragraph{Scheduled Performance} 
On a specific type of GPU server, the performance of LLM inference depends on both the communication patterns of the application task (i.e., regarding to the inference engine mentioned in Section ~\ref{subsec:distributed-llm-serving}) and the hardware topology of resources allocated to LLM instances by resource scheduler \cite{amaral2017topology}. Modern multi-GPU systems have intricate hardware topology, including the connectivity between CPU and CPU, CPU and GPU, GPU and GPU. To ensure efficient execution and maximize resource utilization under co-location, workload schedulers must account for the underlying hardware topology and the communication needs of the workloads when allocating CPU and GPU resources. In what follows, we refer to such performance gain that is significantly affected by resource scheduling as \textit{scheduled performance}.

\paragraph{Illustrative Hardware Topology}
\begin{figure}[!h]
  \centering
  \includegraphics[width=0.75\textwidth]{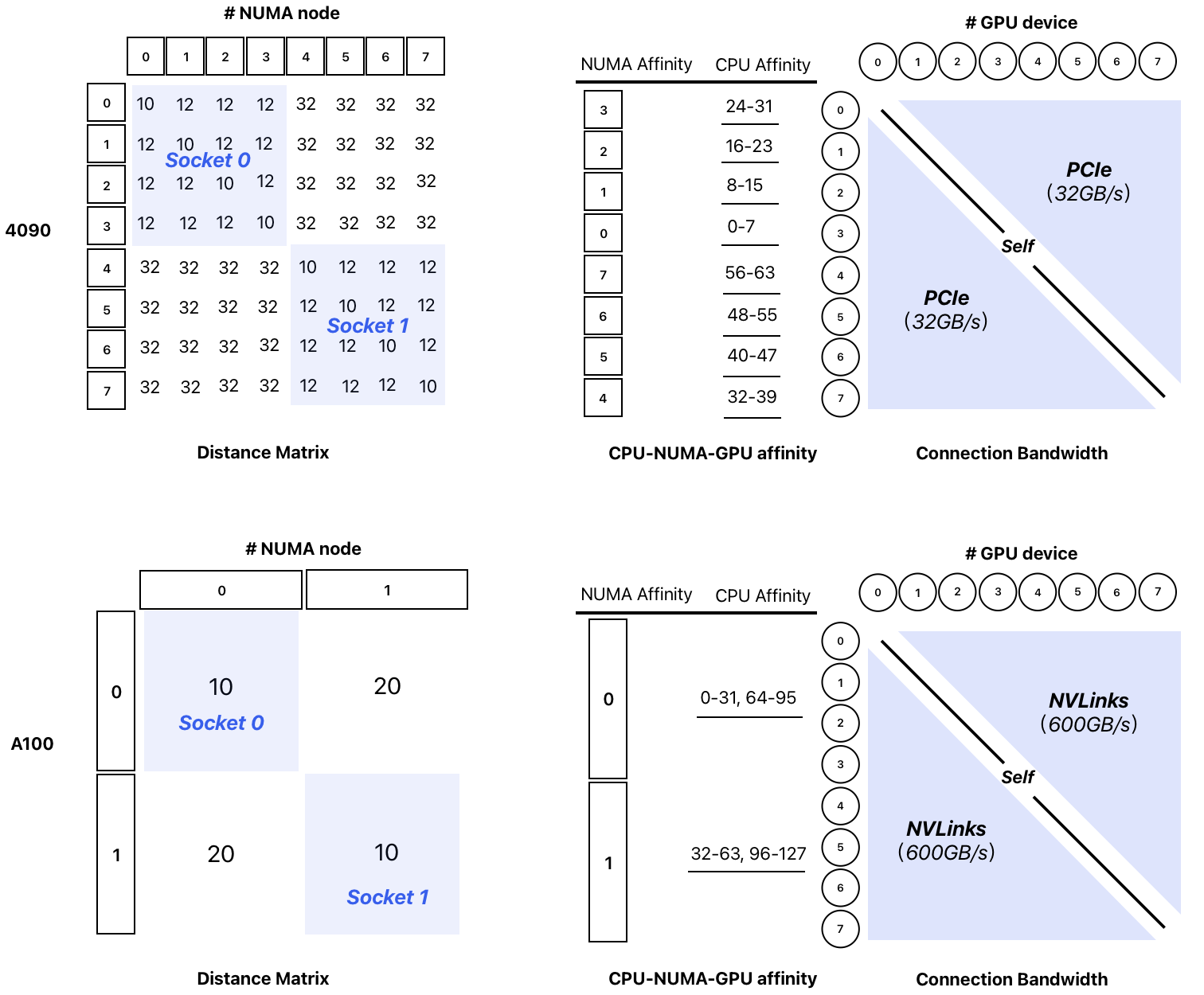}
  \caption{Two hardware topology examples of NVIDIA 4090 server and A100 server. The configurations are 2Sockets-8NUMAs-64Cores \(\times\) 8GPUs of 4090 server, and 2Sockets-2NUMAs-128Cores \(\times\) 8GPUs of A100 server.}
  \label{fig:topo-types}
\end{figure}
To illustrate this issue, Figure ~\ref{fig:topo-types} displays the hardware topology of inference servers equipped with two widely used NVIDIA GPUs, the RTX 4090 and A100, commonly employed in LLM serving. Each of the two servers is equipped with 8 GPUs. The 4090 Server has 64 CPU cores and consists of 2 sockets with each regarding to 4 NUMA nodes \footnote{NUMA is available in multi-CPU systems where each CPU has varying speeds when accessing different memory regions. Memory that is directly attached to a specific CPU is considered “local” to it, enabling faster access, while memory not directly linked is classified as “non-local,” with access times influenced by the number of interconnects between the CPU and the memory. This concept of local versus non-local can also apply to peripheral devices such as NICs and GPUs in modern systems}. 
Figure ~\ref{fig:topo-types} shows that on the 4090 Server the communication between different NUMA nodes within the same socket incurs a \(1.2\)x (\(12\  vs.\ 10\)) higher cost, while communication between NUMA nodes across different sockets is \(3.2\)x (\(32\ vs.\ 10\)) higher, compared to intra-NUMA communication. Respectively, such overhead is \(2\)x (\(20\ vs.\ 10\)) on A100 Server.

Regarding NUMA affinity, ensuring that CPU cores and GPU devices allocated for a LLM instance are aligned within the same NUMA node is essential for maximizing the \textit{scheduled performance}. By keeping both the CPU and GPU local to the same node, memory access latency is minimized, which can significantly improve the efficiency of inference tasks. For example, on the 4090 Server, CPU cores numbered \(24-31\) have their local memory located on NUMA node 3, with the nearest GPU being GPU 0 (In practice, GPU devices are identified by their UUIDs).

\subsection{Auto-scaling of Co-located Workloads}
\label{subsec:auto-scaling-of-co-located-workloads}
It is feasible to guarantee the topology affinity of performance-sensitive workloads during the initial deployment stage. For example, the topology manager \cite{Kubernetes2020topologymanager} of Kubernetes, functioning as a per-node agent (i.e., the \textit{kubelet}), can prevent an instance from starting if, despite sufficient overall resources on the machine, the resource distribution does not align with the required topology policy. However, this issue arises during the \textit{kubelet} admission phase, meaning the scheduler has already made an incorrect scheduling decision by that point. This leads to either the instance failing to start or causing significant performance degradation for high-priority services in co-location scenarios, particularly during auto-scaling and priority-based preemption.

\begin{figure}[]
  \centering
  \includegraphics[width=0.75\textwidth]{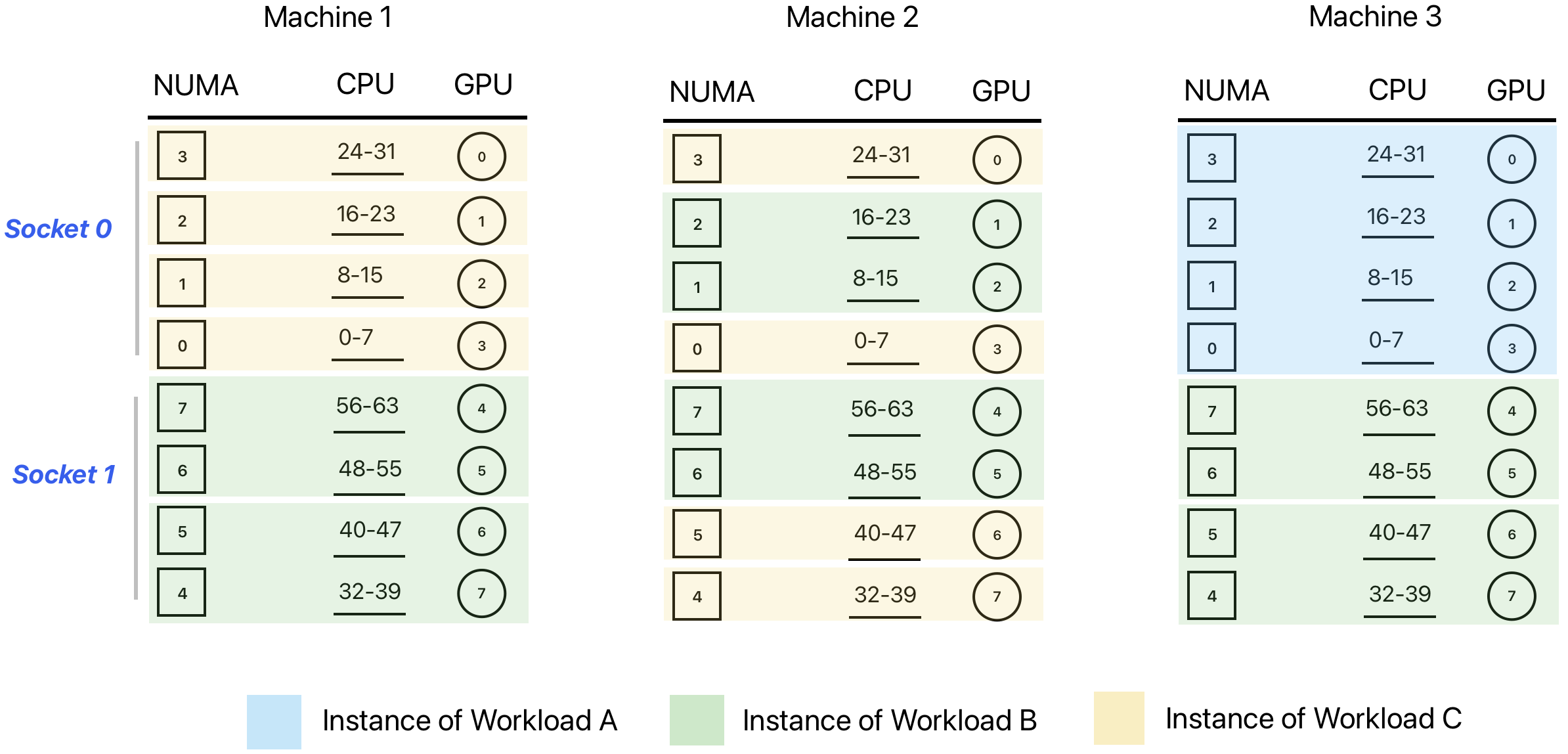}
  \caption{A snapshot of resource allocation in a cluster of 4090 Server for three co-located workloads}
  \label{fig:alloc-snapshot}
\end{figure}

As an illustrative example, Figure ~\ref{fig:alloc-snapshot} presents a snapshot of resource allocation in a cluster of 4090 Server for three co-located workloads. \textbf{Workload A} and \textbf{Workload B} are high-priority, latency-sensitive online inference services with the same priority. They both have strict topology affinity requirements including guaranteed NUMA alignment and best-effort socket-affinity. These workloads cannot be preempted. In contrast, \textbf{Workload C} is a low-priority offline inference job, utilizing cluster resources on a best-effort basis and subject to preemption at any time. Table ~\ref{tab:workload-configuration} summarizes the detailed configuration. The instance specifications of Workload A, B and C are (32cores, 4GPUs), (16cores, 2GPUs), and (8cores, 1GPU), respectively.

\begin{table}[b!]
\captionsetup{skip=5pt} 
\caption{Workload configuration for co-location demonstration}
\centering
\renewcommand{\arraystretch}{1.5} 
\begin{tabular}{l|r|r|r|cc|rr}
\toprule
\multirow{2}{*}{Workload} & \multirow{2}{*}{Priority} & \multirow{2}{*}{Critical} & \multirow{2}{*}{Preemptible} & \multicolumn{2}{c|}{Resource Request} & \multicolumn{2}{c}{Topology Affinity} \\ \cline{5-8} 
    &        &       &       & CPU (cores) & GPU (\#)       & NUMA      & Socket   \\ \hline
A     &   high     &    yes     &   no      &      32       &  4      &   Guaranteed        &     Best-effort   \\ 
B     &   high     &    yes     &   no      &      16       &  2      &   Guaranteed        &     Best-effort   \\ 
C     &   low      &    no      &   yes     &      8        &  1      &   N/A               &     N/A   \\ 
\bottomrule
\end{tabular}
\label{tab:workload-configuration}
\end{table}

In Figure ~\ref{fig:alloc-snapshot}, the snapshot displays one instance of Workload A, six of Workload B, and eight of Workload C scheduled across three machines, all of which are fully allocated, especially in terms of GPU resources. When the demand for high-priority Workload A (or B) spikes, the scheduler initiates preemption to scale up Workload A. At this point, the scheduler must choose preemption candidates from Machine 1 or Machine 2, both of which satisfy the resource requirements. However, selecting Machine 2 would violate the topology affinity of Workload A, leading to resource fragmentation or degraded performance, depending on whether a guaranteed or best-effort affinity policy is used. The core issue here is the scheduler’s lack of topology awareness during preemption. Similarly, if Workload B scales up and triggers preemption, there are 12 possible preemption options (six on each machine), but four of these would fail to meet Workload B’s topology affinity needs.

\section{Topology-aware Preemption}
\label{sec:topology-aware-preemption}
Topology-aware scheduling has been thoroughly studied in existing research. We provide an in-depth review and discussion in Section ~\ref{sec:related-work}. However, the integration of real-time resource topology into preemption decisions between co-located workloads has been largely overlooked. In this section, we present the design of our proposed topology-aware preemption approach. We begin by outlining the overall architecture in Section ~\ref{subsec:overview}, followed by the propose of a unified representation for resource topology of GPU servers in Section ~\ref{subsec:topology-representation} and its maintenance in Section ~\ref{subsec:maintenance-of-topology-information}. Lastly, we introduce the proposed preemption algorithm in Section ~\ref{subsec:algorithm}, offering a thorough analysis of its complexity and the optimizations we have applied.

\subsection{Overview}
\label{subsec:overview}
\begin{figure}[!b]
  \centering
  \includegraphics[width=0.75\textwidth]{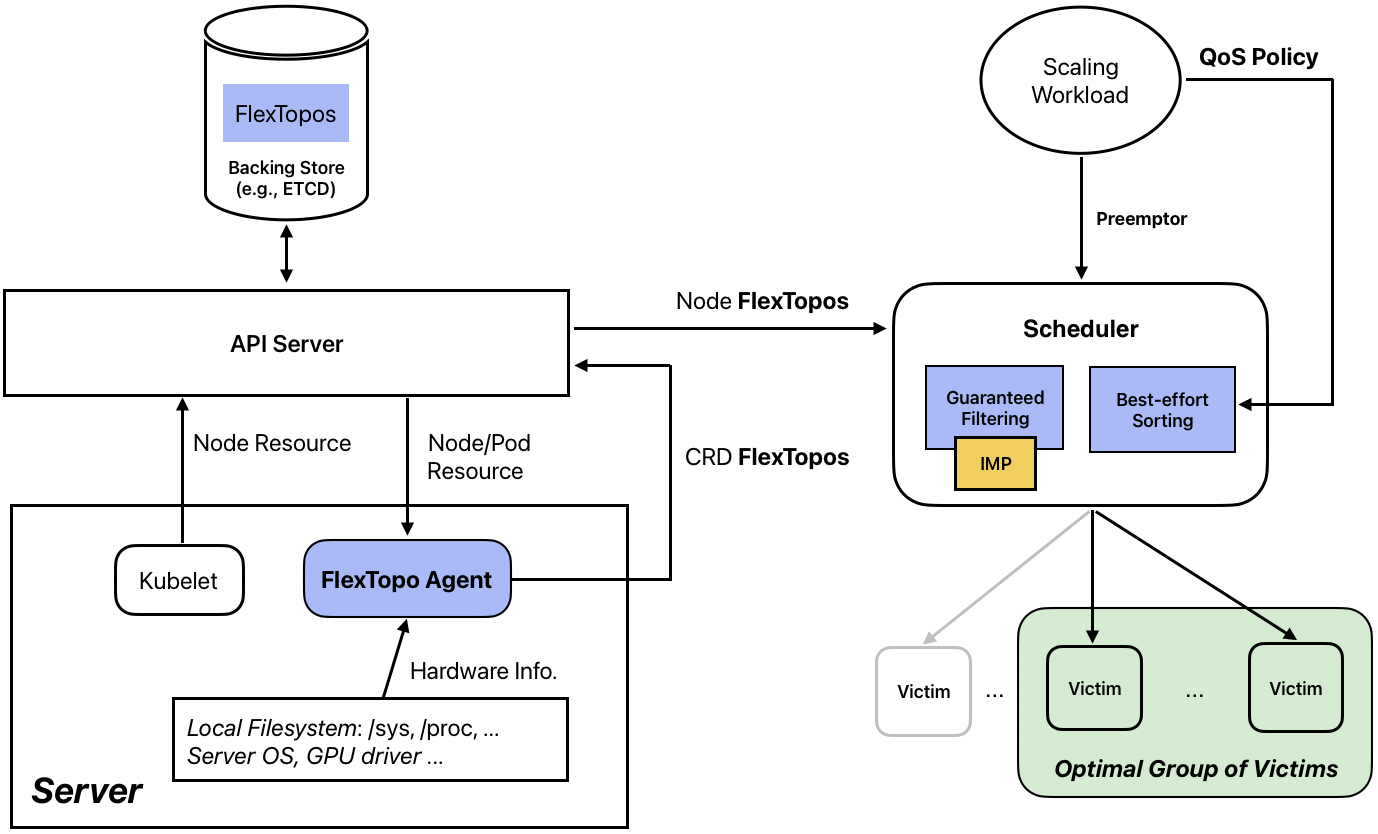}
  \caption{Overview architecture of topology-aware preemption}
  \label{fig:overview-arch}
\end{figure}

To maximize the resource utilization, the scheduling strategy employs \textbf{saturation allocation} during the initial deploy phase, where all available cluster resources are fully allocated by co-locating multiple workloads. When certain tasks experience increased load and require scaling, lower-priority tasks are preempted to free up resources for the expansion. This approach ensures that the cluster operates at maximum capacity, while dynamically adjusting resources through preemption to support scaling demands. We advocate saturation allocation due to the expensive GPU resources and the unacceptable machine-level elasticity overhead, which in practice usually takes half-an-hour, or even much time, to add a new GPU server into the serving cluster when workload traffic spikes. Figure ~\ref{fig:overview-arch} shows the overview architecture of our topology-aware preemption for co-located LLM workloads.

Kubernetes is the standard open-source orchestration system for containerized applications, making it central to our management of distributed LLM serving in production. Accordingly, this paper focuses on resource scheduling within the Kubernetes ecosystem. As shown in Figure ~\ref{fig:overview-arch}, the system architecture consists of three key components: a Custom Resource Definition (CRD) named FlexTopo, the FlexTopo agent responsible for its maintenance, and a topology-aware scheduler featuring Guaranteed Filtering and Best-effort Sorting.

The default Kubernetes scheduler lacks topology awareness, which can lead to Pod startup failures even when a node has sufficient total resources. This occurs because the scheduler may select a node where the resource distribution does not satisfy topology policies during the Kubelet admission phase. A more effective approach would involve informing the scheduler with topology information about available resources on each node, allowing it to choose nodes and victims that satisfy both resource requests and topology constraints of the preemptor. To address this, we propose FlexTopo, a unified representation of real-time resource topology for each server in the cluster. For example, FlexTopo provides the scheduler with detailed information such as Pod A being located on Server 1, with resources allocated on Socket 0, NUMA node 1, CPU cores ranging from 2 to 7, and GPU-0, among others.

The FlexTopo agent, deployed on each node as a daemon, maintains the server’s FlexTopo by gathering hardware topology data from the local file system, server OS, and GPU driver et al., covering components like sockets, NUMA nodes, CPU cores, and GPU devices. This data is combined with resource allocation information from the API server to create a real-time view of node resources, including topology details, known as FlexTopo. When scaling high-priority workloads and initiating preemption, the scheduler employs topology-aware preemption strategies—Guaranteed Filtering or Best-effort Sorting, depending on the preemptor’s QoS requirements, to identify the optimal group of victims.

\subsection{Unified Topology Representation}
\label{subsec:topology-representation}
In terms of hardware topology, current research often relies on specific hierarchical models and lacks a standardized representation that can universally describe all types of mainstream GPU servers, including those using NVIDIA devices. This limitation makes topology-aware scheduling methods highly specific to certain setups. As a result, their generalizability is reduced, leading to increased deployment costs due to the need for adaptation in each use case. 

To accommodate the diversity in hardware architectures, we propose \textbf{FlexTopo} (i.e., \textbf{Flex}ible \textbf{Topo}logy), a unified resource topology representation that employs a flexible, graph-based model to accurately depict both static hardware topology and dynamic resource allocation states. In FlexTopo, hardware components, such as sockets, CPU cores, NUMA nodes, and GPUs, are represented as nodes, while edges depict the physical or logical topology between them. Each node and edge is annotated with attributes, as summarized in Table ~\ref{tab:flexTopo}. Note that in FlexTopo we omit the details of PCIe switches, NVSwitches et al. aimed to only focusing on informed scheduling decisions.

\begin{table}[!b]
\captionsetup{skip=5pt} 
\caption{Node and edge definitions in FlexTopo for informed preemption scheduling}
\centering
\renewcommand{\arraystretch}{1.4} 
\begin{tabular}{l|l|l|ll}
\toprule
\textbf{Components} & \textbf{Component Type} & \textbf{Attributes}  & \textbf{Description} \\ \hline
\multirow{2}{*}{Node} 
& Socket             & Socket ID                    & -                     \\ 
&                    & Extensible                   &                       \\ \cline{2-4} 
\multirow{3}{*}{} 
& CPU CoreGroup      & CoreGroup ID                 & - \\ 
&                    & Status                 & free/allocated \\ 
&                    & UsedBy                 & instance/pod name                 \\ \cline{2-4} 
\multirow{3}{*}{} 
& CPU Core      & Core ID                 & -                     \\ 
&               & Status                  & free/allocated        \\ \cline{2-4}
\multirow{4}{*}{} 
& NUMA Node     & NUMA node ID             & -  \\ 
&               & Extensible           &    \\ \cline{2-4} 
\multirow{4}{*}{} 
& GPU Device    & UUID                     & GPU ID \\ 
&               & Model                    & e.g., NVIDIA RTX 4090, SXM A100\\ 
&               & Memory Capacity          & VRAM in megabyte \\ 
&               & Status                   & free/allocated \\ 
&               & UsedBy                   & instance/pod name  \\ \hline
\multirow{2}{*}{Edge} 
& Socket \textemdash{} CPU CoreGroup & Host  & Socket \textit{hosts} CPU CoreGroup \\ 
\multirow{3}{*}{} 
& CoreGroup \textemdash{} CPU core & Contain    & CoreGroup \textit{contains} multiple cores \\ 
\multirow{3}{*}{} 
& CoreGroup \textemdash{} NUMA node & Localized    & CoreGroup is \textit{localized} to NUMA \\ 
\multirow{3}{*}{} 
& GPU \textemdash{} NUMA node & Nearby             & GPU is \textit{nearby} NUMA \\ 
\bottomrule
\end{tabular}
\label{tab:flexTopo}
\end{table}

As shown in Table~\ref{tab:flexTopo}, we group multiple CPU cores into a unit called a CPU CoreGroup. This approach is motivated by two main factors. First, modern GPU server used for LLM serving often come with a large number of CPU cores—such as SXM A100 servers, which may have hundreds of cores. Constructing a FlexTopo graph at the individual core level would result in a graph that is too complex to process efficiently. Second, in practice, GPU-based workloads, especially LLM applications, typically request more than just one or two CPU cores. Grouping CPU cores simplifies the FlexTopo graph and aligns with common allocation practices, making it more efficient to process and manage. For example, widely used inference servers like the 4090 and A100 often come with 8 GPU devices, where GPU resources are typically the dominant resource. Grouping cores, such as bundling 8 cores with a GPU device, matches practical allocation patterns. Furthermore, to accommodate different hardware architectures and varying workload requirements, FlexTopo allows the CoreGroup size to be configurable.

In FlexTopo’s edge definition, we specify four types of connections between nodes: \textit{host, contain, localized}, and \textit{nearby}. For instance, one part of the hardware topology of a server can be described as follows: Socket 0 hosts CPU CoreGroup 0, which contains CPU cores 2-7. Each of these cores is localized to NUMA node 0, while GPU 0 is nearby NUMA node 0. These connection types capture both the hierarchical and proximity relationships within the server’s hardware.

\begin{figure}[!h]
  \centering
  \includegraphics[width=0.95\textwidth]{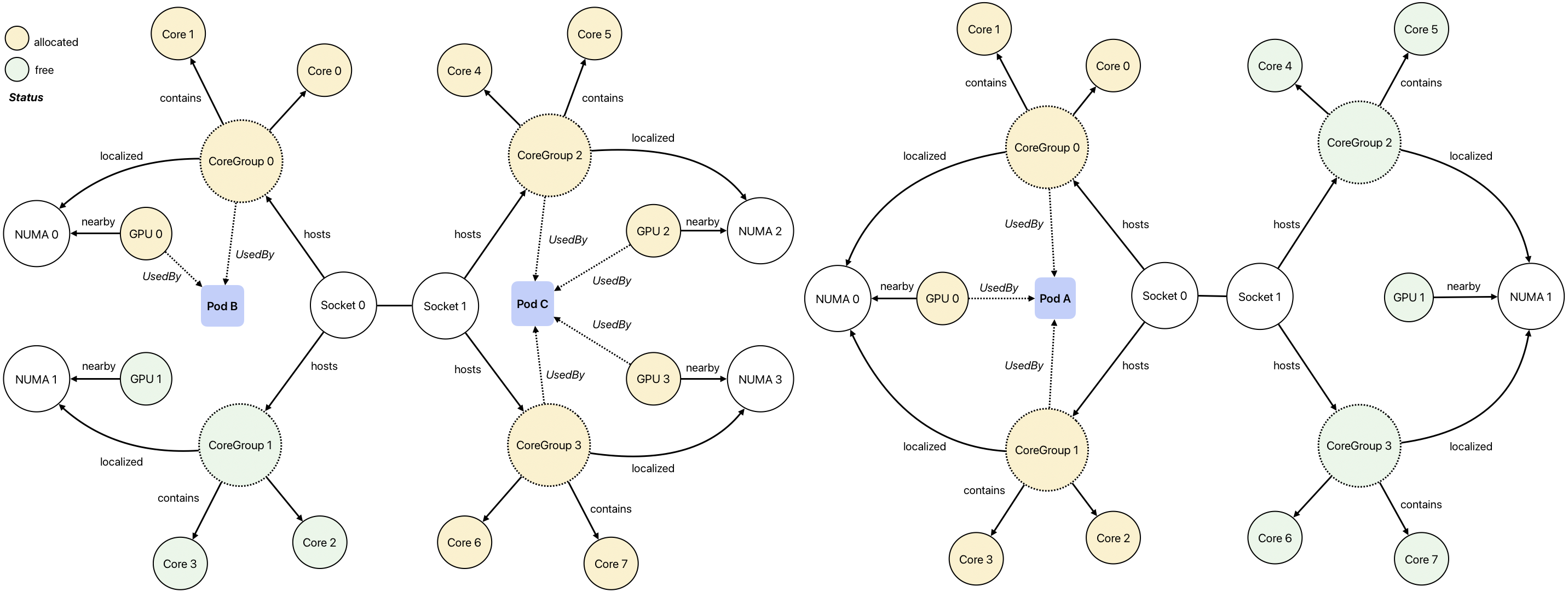}
  \caption{Two illustrative examples of FlexTopo and resource allocation states}
  \label{fig:flextopo-examples}
\end{figure}

To integrate dynamic resource allocation states into the server’s hardware topology, we use the node attributes \textit{Status} and \textit{UsedBy} to establish a logical reference between FlexTopo nodes and the scheduled instances or pods. These attributes provide information about the current state and allocation of resources. Figure ~\ref{fig:flextopo-examples} illustrates two examples of FlexTopo configurations, each demonstrating different hardware topology and resource allocation scenarios \footnote{The FlexTopo examples provided here are simplified, featuring only a few nodes to ensure clear and straightforward demonstration, highlighting key concepts without overwhelming complexity.}. The diagrams demonstrate how the \textit{Status} and \textit{UsedBy} attributes capture the real-time mapping between hardware components and active workloads. Meanwhile, they offer an overview of FlexTopo’s flexibility as a unified topology representation.

\subsection{Maintenance of Topology Information}
\label{subsec:maintenance-of-topology-information}
To maintain the FlexTopo for each server in the cluster efficiently but in a timely manner, practical considerations must be accounted for in the design of the FlexTopo agent. The motivation is to ensure that hardware topology and resource allocation information remain up-to-date, while also minimizing system overhead and avoiding unnecessary strain on the underlying infrastructure.

The hardware topology of a server may change under two primary conditions: (1) a complete server failure, and (2) the failure of one or more GPU devices. In the first scenario, the scheduler and Kubernetes itself handle this issue by detecting the absence of status updates from the Kubelet or by identifying a failed node through a health check mechanism. This ensures that the rest of the cluster is aware of the server failure and can redistribute workloads as needed. In the second scenario, topology changes due to the failure of GPU devices \textit{nearby} specific NUMA nodes, which can affect preemption decision.

However, hardware failures are relatively infrequent in modern GPU servers, and while such failures can be significant in large-scale clusters, the frequency of these events for individual servers tends to be low. Therefore, we adopt a periodic check of the hardware topology on each server. After each check, the results are compared against the internally maintained hardware state. If discrepancies are detected, FlexTopos are updated to reflect the newest hardware configuration. 

In contrast, resource allocation states tend to fluctuate more frequently, particularly in large-scale clusters where many diverse workloads are continuously scheduled, running, and terminated. Timely updates to the FlexTopo CRD reflecting changes in resource allocation are critical. However, frequent updates can strain the API server and ETCD. To mitigate this, instead of continuously polling and updating, the agent updates its FlexTopo using an event-driven approach. It reports updates to the FlexTopo CRD only when changes in resource allocation are detected, minimizing unnecessary strain on the cluster’s control plane. By leveraging infrequent periodic checks for hardware stability and timely event-driven updates for resource allocation state, the agent maintains a high degree of FlexTopo accuracy and operational efficiency.

\subsection{Preemptive Scheduling}
\label{subsec:algorithm}
\paragraph{Pipeline of Preemptive Scheduling}
Using FlexTopo, the scheduler can make topology-aware preemption decisions. Algorithm~\ref{alg:topology-aware-preemption-pipeline} outlines the primary pipeline in this preemptive scheduling process. Since preemptors often have specific topology requirements, as shown in Table~\ref{tab:workload-configuration}, we distinguish two scenarios of topological QoS: \textit{Guaranteed} and \textit{Best-Effort}.

\SetKwComment{Comment}{/* }{ */}
\begin{algorithm}[hbt!]
\setstretch{1.3}
\caption{Pipeline of Topology-aware Preemptive Scheduling}
\label{alg:topology-aware-preemption-pipeline}
\KwData{preemptor $P$, candidate set with victims $\{(N, V_N)\}$}
\KwResult{the optimal candidate $(N^*, v^*_{N^*})$}
\If{$P.\text{qos} == \textit{guaranteed}$}{
    Filtering($N.\text{flextopo}$, $V_N$) $\to$ Sorting($N.\text{flextopo}$, $V_N$)\;
}
\uElseIf{$P.\text{qos} == \textit{best-effort}$}{
    Filtering($N, V_N$) $\to$ Sorting($N.\text{flextopo}$, $V_N$) \Comment*[r]{No enforced filtering for best-effort}
}
\Else{
    Filtering($N, V_N$) $\to$ Sorting($N, V_N$)\;
}
\end{algorithm}

For \textit{Guaranteed} QoS, topology evaluation begins at the \textit{Filtering} stage. Here, each candidate node and its potential victims are assessed under a hypothetical scenario where all victims are initially supposed to be drained to maximize available resources. If, after this process, the preemptor’s requirements are met, the candidate is deemed feasible. The \textit{Filtering} stage focuses only on candidate feasibility and not on specific victim selection, which is deferred to the \textit{Sorting} stage. For \textit{Best-Effort} QoS, no topology constraints are enforced during \textit{Filtering}, allowing as many candidates as possible to pass through. This maximizes the likelihood of identifying a globally optimal preemption solution.

In the \textit{Sorting} stage, each candidate’s topology fitness is evaluated using its state as described by FlexTopo, with a piecewise linear scoring function. Candidates that support allocation and alignment within the same NUMA node receive the highest score. If only the same socket is aligned and achievable, a moderate score is assigned, while candidates requiring cross-socket allocation receive the lowest score. This scoring mechanism allows for a ranked comparison of candidates’ fitness, in terms of topology affinity, for the preemptor.

\begin{figure}[h]
  \centering
  \includegraphics[width=0.95\textwidth]{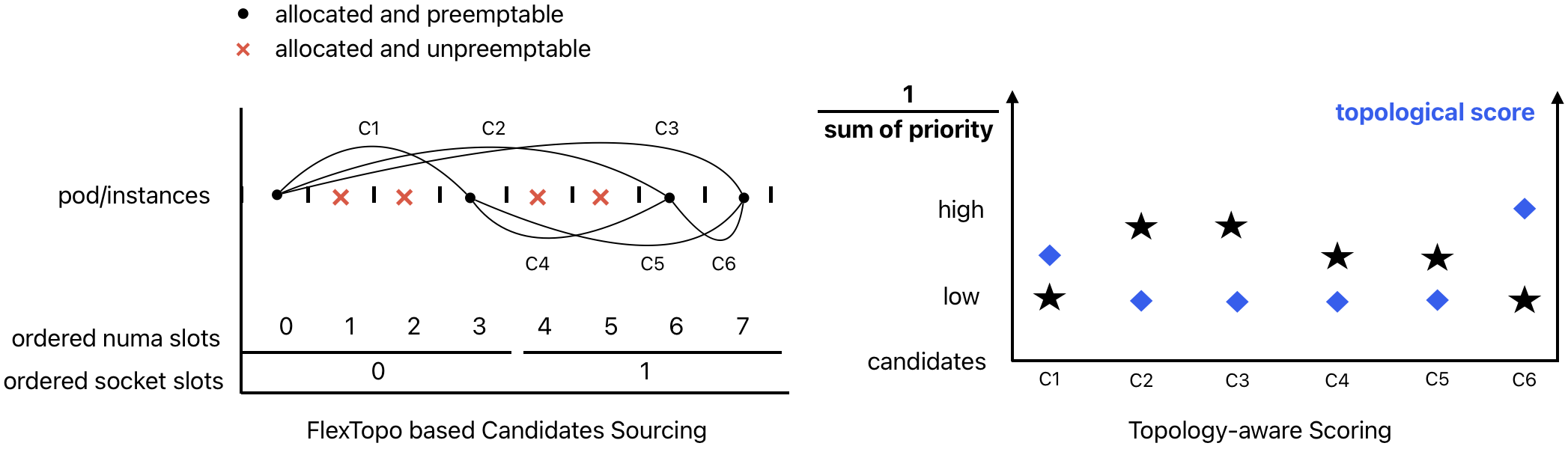}
  \caption{Topology-aware candidates sourcing and scoring}
  \label{fig:score-function}
\end{figure}

\paragraph{Candidates Sourcing and Scoring}
Unlike the default Kubernetes scheduler, in the \textit{Sorting} stage, we treat each candidate as a unique node with varying sets of potential victims. This means that a single candidate node may offer multiple feasible preemption solutions. For example, on a candidate node, preemption may involve selecting different combinations of low-priority pods (victims) to free up the required resources. Each combination offers a distinct solution with varying topology characteristics, such as NUMA alignment or socket locality, which impacts the preemptor’s performance. By evaluating these combinations, we identify the solution that best meets the preemptor’s topology preferences, minimizing cross-socket/NUMA resource allocation when possible.

The widely used preemption primarily focuses on selecting victims based on priority \cite{preemption2024k8s}, aiming to minimize individual or total priority among candidates. In this paper, we extend this approach by incorporating topological considerations when sourcing preemption candidates. Figure~\ref{fig:score-function} illustrates how both priority and the topological distribution of victims are evaluated concurrently. Specifically, we define two sub-scores within our scoring function: a priority-based score, which inversely relates to the total priority, and a topological score, reflecting the distribution of selected victims' resources across the node. The two sub-scores are combined using a weighted linear function as shown by Equation~\ref{eq:score-func}.

\begin{equation}
\label{eq:score-func}
    S(C) = \alpha \times \frac{1}{\text{sum of priority(C)}} + (1-\alpha) \times T(C_{\text{flextopo}})
\end{equation}

In Equation~\ref{eq:score-func}, $C=(N, v_N) \in \{(N, v_N) | v_N \in V_N\}$ represents a candidate (a node with a set of potential victims on it); $T$ is the topological score, given by

\[
T(C_{\text{flextopo}}) = 
\begin{cases} 
\text{high}, & \text{if } v_N \in \text{ the same aligned NUMA of } N  \\
\text{medium}, & \text{if } v_N \in \text{ the same aligned socket but different NUMA of } N \\
\text{low}, & \text{if } v_N \text{ is located across sockets}
\end{cases}
\]

To select the optimal candidates, we maximize $S$. The parameter $\alpha \in [0, 1]$ controls the balance between priority and topology affinity; setting $\alpha = 0$ indicates a preference purely for resource topology, while $\alpha = 1$ emphasizes priority alone. By identifying the best candidate within each node and then comparing the top candidates across nodes, we search the globally optimal selections. With the scoring function defined by Equation~\ref{eq:score-func}, the \textit{Sorting} stage of preemption can be formulated by Equation~\ref{eq:score-object}.

\begin{equation}
\label{eq:score-object}
(N^*, v^*_{N^*}) = \mathop{\arg\max}_{C=(N, v_N), v_N \in V_N} S(C)
\end{equation}

\paragraph{Incremental Minimal Preemption} 
To maximize Equation~\ref{eq:score-object}, a straightforward approach involves generating all possible subsets of potential victims—that is, the pods eligible for preemption—and validating each subset on every candidate node to determine if removing those pods would allow the preemptor to be scheduled. This approach, however, has exponential time complexity. Specifically, if there are  $m$  potential victims on a candidate node, the total number of non-empty subsets to evaluate is $\sum_{i=1}^{m} C(m, i) = 2^m - 1 = O(2^m)$. Even with parallel processing to expedite the evaluation of all combinations, the computational overhead may remain unacceptable, particularly in large-scale clusters where scheduling throughput is a critical metric. 

\SetKwComment{Comment}{/* }{ */}
\begin{algorithm}[hbt!]
\setstretch{1.3}
\caption{Incremental Minimal Preemption}
\label{alg:incremental-minimal-preemption}
\KwData{$P$, $N$, a set of potential victims $V_N$=$\{v_1, v_2, ..., v_m\}$}
\KwResult{$R$, a group of feasible subsets of victims at size $k, k \le m$}
$R \gets \emptyset $\;
\If{$\mathcal{S}(P, N\!\setminus\!V_N) = 0$}{ 
    Return $R$  \Comment*[r]{Extreme validation for failfast}
}
$k \gets 1$\;
\While{$k \le m$}{
  $C \gets \{ s \subset V_N \mid |s| = k \text{ and } s \neq \emptyset \}$ \;
  \For{$s \in C$}{
      \If{$\mathcal{S}(P, N\!\setminus\!s) = 1$}{
         $R \gets R \cup s$
      }
  }
  \If{$|R| \ge 1$}{
      Return $R$     \Comment*[r]{Minimal found, early stopping}
  }{
      $k \gets k + 1$\;
  }
}
\end{algorithm}

To address this challenge, we propose an algorithm called Incremental Minimal Preemption (IMP), given by Algorithm~\ref{alg:incremental-minimal-preemption}. In Algorithm~\ref{alg:incremental-minimal-preemption}, $\mathcal{S}: P \times N \to \{0, 1\}$ is an indicator function of scheduling, where $\mathcal{S}(P, N) = 1$ represents that instance $P$ can be scheduled on node $N$ while $\mathcal{S}(P, N) = 0$ means can not; $N\!\setminus\!V$ means evicting all the victim pods in $V$ from node $N$, hypothetically; $|s|$ denotes the cardinality of set $s$.

IMP employs a greedy strategy that aims to reprieve as many victims as possible during preemption. The key idea is to evaluate subsets of victims starting from the smallest subset size (e.g., size = $1$). Once feasible groups of victims are found at the current subset size, the algorithm stops searching larger subset sizes. For instance, if the algorithm identifies feasible victim groups consisting of two pods each (i.e., size = $2$), there is no benefit in considering the removal of three or more pods (i.e., size $\ge 3$), since the preemptor can be scheduled after removing just two victims. This incremental approach often terminates the search early and reduces overall computational overhead, allowing us to find the minimal group of victims efficiently, as detailed in the bellow.

\paragraph{Complexity Analysis of IMP} For a given combination size $k$, the number of possible combinations is $C(m, k) = \frac{m!}{k!(m - k)!}$. Thus, the total number of combinations to consider up to size  $k$ is $\sum_{i=1}^{k} C(m, i)$, where $k$ is the minimal combination size that allows the preemptor to be scheduled. In the \textbf{worst-case scenario}, if no combinations satisfy the scheduling condition until  $k = m$, we need to traverse all possible combinations. The total number of combinations in this case is  $\sum_{i=1}^{m} C(m, i) = 2^m - 1$, which means the time complexity degrades to exponential,  $O(2^m)$, similar to the straightforward approach.

However, in the \textbf{average case}, practical preemptors, especially GPU workloads like LLM applications, typically require removing only a small number of pods (i.e., $k$ is usually small) for successful scheduling. Assuming feasible combinations are found when $k$ is small, the total number of combinations becomes $\sum_{i=1}^{k} C(m, i)$. When $k$ is small—for example,  $k = 1$  or $k = 2$ — the number of combinations reduces significantly. Specifically, when $k = 1$, the number of combinations is $C(m, 1) = m$; when $k = 2$, it is $C(m, 1) + C(m, 2) = m + \frac{m(m - 1)}{2}$. Therefore, on average, the IMP achieves a time complexity that is close to polynomial rather than exponential.

\section{Implementation}
\label{sec:implementation}
We present the implementation of the proposed topology-aware preemption in three key components: the FlexTopo API, the FlexTopo Agent, and the Scheduler Plugin. The FlexTopo API and Agent code are available at \url{https://github.com/agiping/flextopo}, and the Scheduler Plugin code can be found at \url{https://github.com/agiping/godel-scheduler}.

\begin{figure}[!h]
    \centering
    \includegraphics[width=0.6\linewidth]{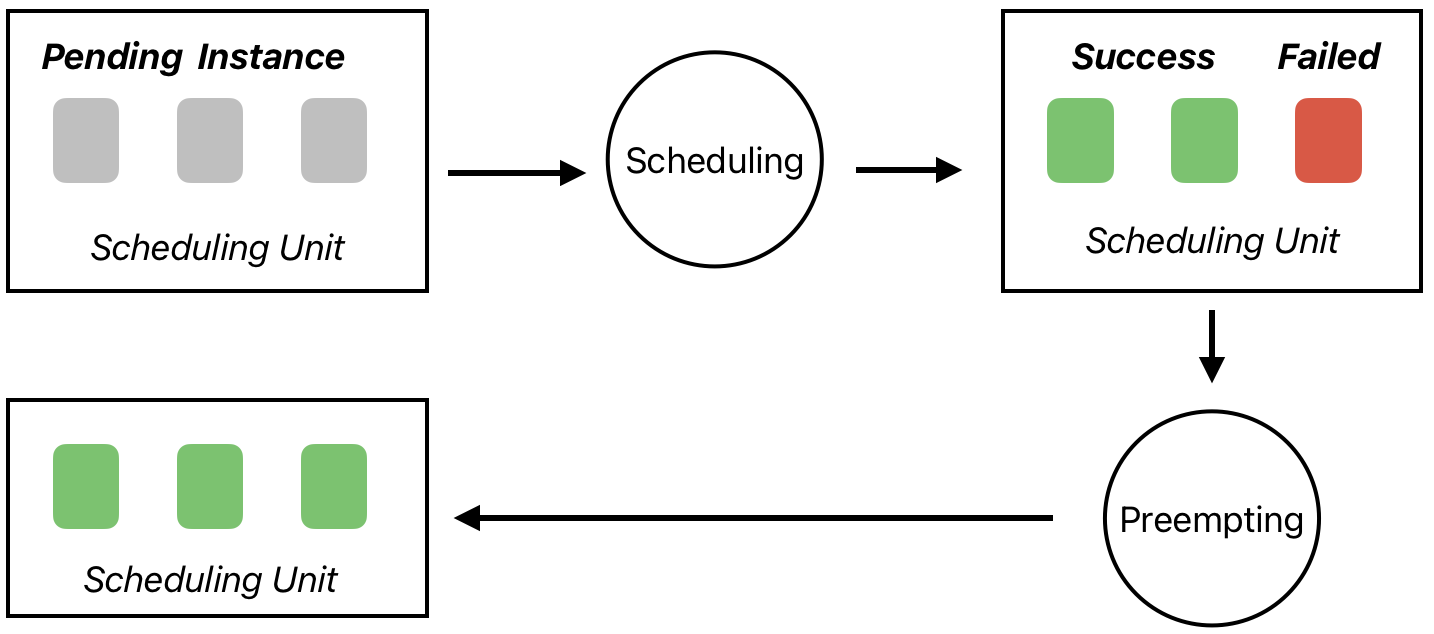}
    \caption{Scheduling and preemption workflow in Gödel\cite{kubewharfGodel-scheduler}}
    \label{fig:workflow-of-Godel}
\end{figure}

\textbf{FlexTopo API.} To decouple data collection from scheduling logic, we developed the FlexTopo API as an independent, shared module utilized by both the FlexTopo agent and the Scheduler. This API defines a lightweight FlexTopo CRD structure and offers a comprehensive client set for both the Agent and Scheduler, enabling interaction with the API server for reading and writing FlexTopo data. By isolating the API as a standalone component, we ensure code consistency between the Agent and Scheduler, aligning with best practices of Kubernetes community.

\textbf{FlexTopo Agent.} The FlexTopo agent operates as a DaemonSet \cite{daemonset} on each node. Its implementation adheres to the core design principles discussed in Section~\ref{subsec:maintenance-of-topology-information}, focusing on accuracy and performance of data collection. Notably, due to the need for collecting hardware-level topology data, the agent requires privileged security permissions, which is essential for the implementation.

\textbf{Scheduler Plugin.} The preemptive scheduling policy is implemented as a plugin within the Gödel Scheduler \cite{kubewharfGodel-scheduler}, a robust scheduling and resource management platform proposed and open-sourced by ByteDance. Gödel is designed for cloud-native workloads, it enhances the default Kubernetes scheduler with support for both online and offline scheduling. The high-level scheduling and preemption workflow of Gödel is depicted in Figure~\ref{fig:workflow-of-Godel}. By leveraging the abstract concept of a \textit{Scheduling Unit}, Gödel unifies independent instance scheduling and Gang scheduling under a cohesive framework. For each scheduling task, instances are initially scheduled through the standard scheduling cycle. Instances that fail this cycle are subsequently processed through the preemption mechanism.

\section{Evaluation}
\label{sec:evaluation}
We evaluate the proposed topology-aware preemption through both simulations and near-production clusters. Due to the risks of extensive testing under high load in production—potentially disrupting regular workloads—we primarily rely on simulations to validate our approach. In production, we assess resource allocation distributions before and after deploying our topology-aware solution, while more detailed analyses use simulated clusters with varied LLM workloads.

\begin{figure}[!b]
  \centering
  \begin{subfigure}[b]{0.75\textwidth}
    \centering
    \includegraphics[width=\textwidth]{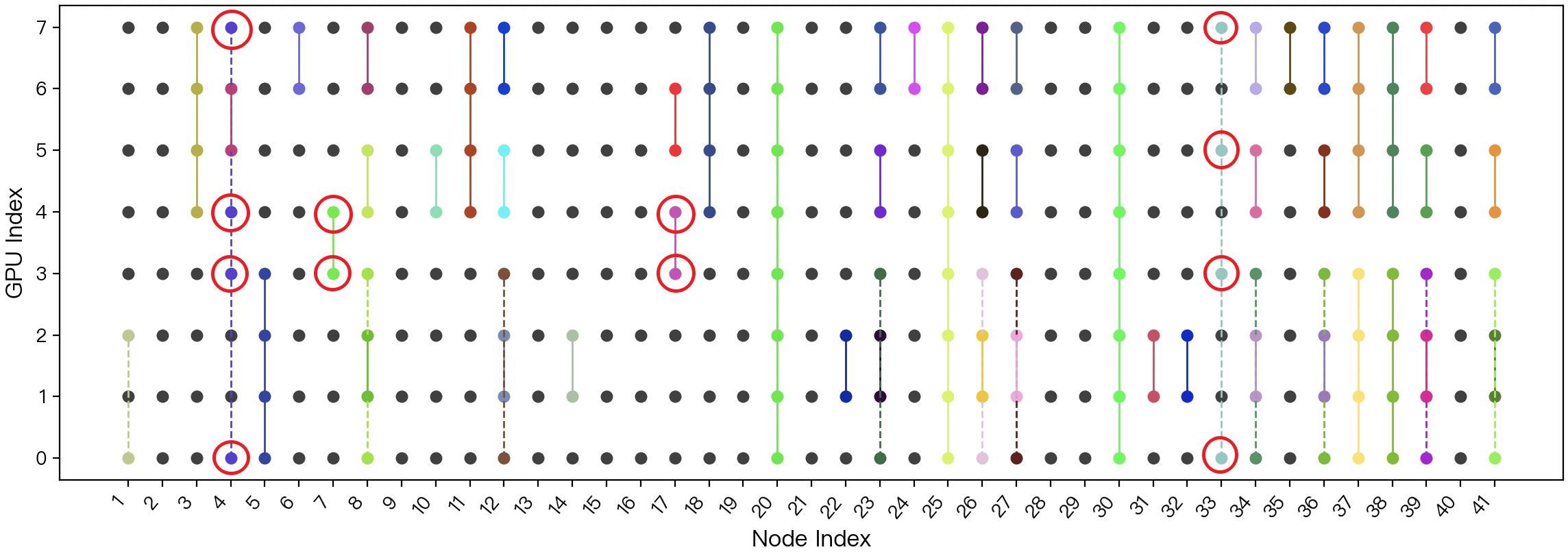}
    \caption{Without topology-aware scheduling}
    \label{fig:resource-allocation-before-sampling}
  \end{subfigure}
  \hfill
  \begin{subfigure}[b]{0.75\textwidth}
    \centering
    \includegraphics[width=\textwidth]{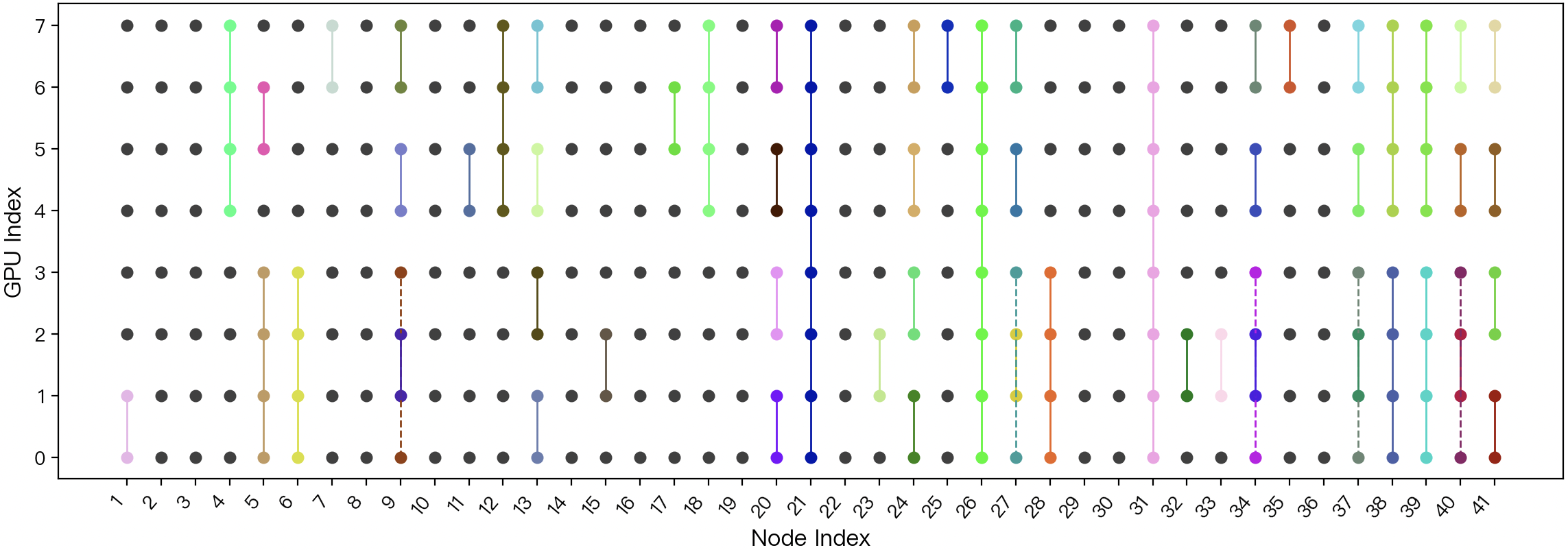}
    \caption{With FlexTopo-based preemption}
    \label{fig:resource-allocation-after-sampling}
  \end{subfigure}
  \caption{Snapshot of GPU allocation distribution before and after deploying topology-aware scheduling}
  \label{fig:resource-allocation-comparison}
\end{figure}

\paragraph{Resource Allocation Distribution} 
To illustrate the topology-aware improvements in scheduling, Figure~\ref{fig:resource-allocation-comparison} presents snapshots of the GPU allocation before and after deploying FlexTopo-based preemption. This cluster, a near-production environment, includes 41 GPU servers, each with 8 NVIDIA RTX 4090 GPUs. Each server has a 2-socket, 8-NUMA hardware configuration. Each GPU is aligned with one NUMA node, of which those indexed with $0 - 3$ are hosted by Socket $0$ and $4 - 7$ are hosted by Socket $1$.

In Figure~\ref{fig:resource-allocation-comparison}, the vertical axis represents ordered GPU device indices, and the horizontal axis denotes servers. Multiple co-located workloads use these GPUs, with allocations ranging from 1, 2, 4, to 8 cards per instance. Lines link GPUs allocated to instances requiring multiple cards, with dotted lines indicating cases where allocations span non-contiguous indices. 

As shown by Figure~\ref{fig:resource-allocation-comparison}(a), several instances have GPU allocations crossing sockets (on 2-socket and 8-NUMA hardware, they are crossing sockets while they are crossing NUMAs on 2-socket and 2-NUMA hardware), which fails to meet the topology affinity requirement for latency-sensitive LLM workloads. For example, instances scheduled on Nodes 4, 7, 17, and 33 (marked by red circles) show this issue. This misallocation may result in either failed scheduling for guaranteed QoS or performance degradation for best-effort. By using FlexTopo-based preemption, these cross-socket (or cross-NUMA) allocations are eliminated.

It is important to note that due to the high dynamism of workload traffic in near-production cluster, the allocation states change continuously with auto-scaling. Therefore, the snapshots in Figure~\ref{fig:resource-allocation-comparison} may not be strictly comparable in terms of specific instances. Nevertheless, the distribution shows a clear improvement in GPU allocation topology, reducing cross-socket issues, ensuring NUMA affinity and improving overall efficiency.

\paragraph{Simulation Configuration of KWOK} To have an in-depth investigation, we simulate a 100-node GPU cluster with type of RTX 4090 using \textit{Kubernetes Without Kubelet (KWOK)} \cite{kwok2024k8s}. KWOK provides a lightweight, fast solution for creating Kubernetes clusters, in which all nodes are simulated to mimic the behavior of actual cluster nodes. 

\begin{table}[!h]
\captionsetup{skip=5pt} 
\caption{Workload configuration for preemption simulation}
\centering
\renewcommand{\arraystretch}{1.2} 
\begin{tabular}{l|l|c|c|c|c}
\toprule
Workload & Priority & \multicolumn{1}{l|}{Type} & \multicolumn{1}{l|}{GPU/Instance} & \multicolumn{1}{l|}{Initial Number of Instancs} & \multicolumn{1}{l}{Can Be Preempted} \\ \hline
A        & 1500     & Online                    & 8                                 & 20                                              & False                                \\
B        & 1000     & Online                    & 4                                 & 40                                              & False                                \\
C        & 500      & Offline                   & 2                                 & 200                                             & True                                 \\
D        & 200      & Offline                   & 1                                 & 80                                             & True                                 \\ 
\bottomrule
\end{tabular}
\label{tab:workload-for-preemption}
\end{table}

\begin{figure}[!h]
    \centering
    \includegraphics[width=0.85\linewidth]{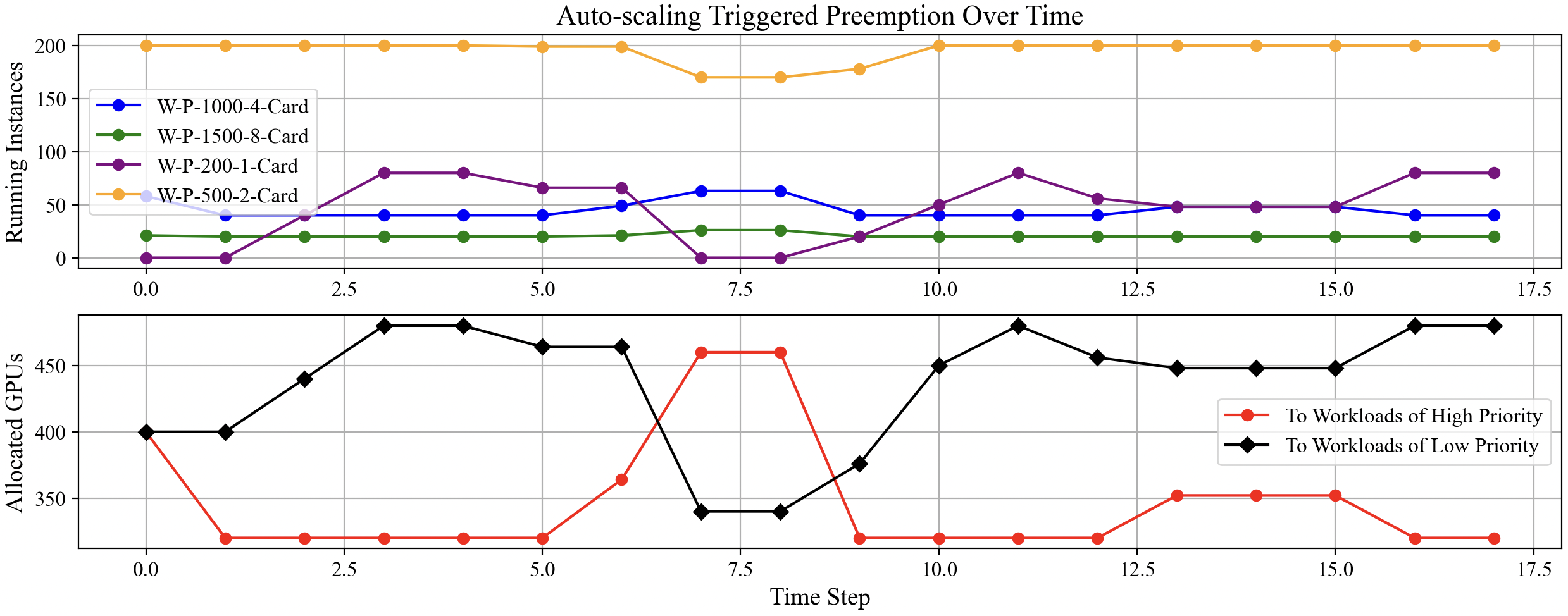}
    \caption{Number of instances during preemption}
    \label{fig:running_replicas}
\end{figure}

Table~\ref{tab:workload-for-preemption} outlines the workload configuration, including four types with varying priorities and resource requirements. Initially, all resources across the 100 nodes are fully allocated, achieving the saturation allocation described in Section~\ref{subsec:overview}. Figure~\ref{fig:running_replicas} shows an illustrative simulation triggered by $2$ auto-scaling events for high-priority workloads.

\paragraph{Topology Affinity Hit Rate} To evaluate the improvements achieved with FlexTopo-based preemption, we analyze the topology affinity hit rate for Workloads B and C (shown in Table~\ref{tab:workload-for-preemption}), where each instance requests 4 and 2 GPU devices, respectively. We compare the hit rate during preemption under Gödel’s standard approach and our topology-aware method. Notably, workloads requiring 8 GPU devices, such as Workload A, are excluded from this hit rate analysis, as their allocation inherently ensures either complete affinity or none at all; with 8 devices, a single server is fully occupied, rendering GPU allocation as the dominant factor. Similarly, Workload D, assigned the lowest priority, is also excluded from the hit rate analysis.

Table~\ref{tab:topo-hit-rate} presents the results. For a robust comparison, we initiated 100 simulation cycles under conditions of saturated allocation, with each cycle triggering 50 instances to scale up. This setup equates to 50 independent preemptions per cycle, meaning that for each instance scaled up, the candidate sourcing and victim selection processes are evaluated independently. Consequently, a total of $100 \times 50 = 5000$ preemptions were assessed.

In these 5000 evaluations, Gödel’s standard preemption policy met the topology requirements of Workloads B and C a total of 2225 times, achieving a hit rate of approximately $45\%$. In contrast, the proposed FlexTopo-based preemption approach satisfied topology requirements in all 5000 cases, resulting in a hit rate of $100\%$. This improvement means that for guaranteed topology QoS, the scheduling failure rate (such as Kubernetes TopologyAffinityError\cite{topoerrork8s}) is reduced by $55\%$. While for workloads with best-effort QoS, FlexTopo-based preemption provides a $55\%$ improvement in \textit{Scheduled Performance}.

\begin{table}[]
\captionsetup{skip=5pt} 
\caption{Topology affinity hit rate of Workload B and C during 100 $\times$ 50 preemptions}
\centering
\renewcommand{\arraystretch}{1.2} 
\begin{tabular}{l|c|r|r}
\toprule
Method                       & No. Preemptions & No. Hit  & Hit Rate \\ \hline
Gödel Standard Preemption    & 100 $\times$ 50          & 2225 & 44.5\%     \\
Gödel + FlexTopo             & 100 $\times$ 50          & 5000 & 100\%    \\
\bottomrule
\end{tabular}
\label{tab:topo-hit-rate}
\end{table}

\paragraph{Overhead Analysis} Scheduling cost is a critical metric for evaluating performance in large-scale clusters. To assess the overhead introduced by topology-aware preemption and our optimization based on IMP~\ref{alg:incremental-minimal-preemption}, we measure the time cost under three conditions: Gödel’s standard preemption, FlexTopo-based preemption without IMP optimization, and FlexTopo-based preemption with IMP optimization. The analysis specifically focuses on the time cost incurred during the candidate sourcing phase within the preemption pipeline, as this is the primary contributor to time overhead in our observation. The scheduler is allocated a configuration of 1 CPU and 1 GiB of memory (1U-1Gi) and runs on a Linux server with AMD/64-bit CPU architecture.

\begin{figure}[!b]
    \centering
    \includegraphics[width=0.88\linewidth]{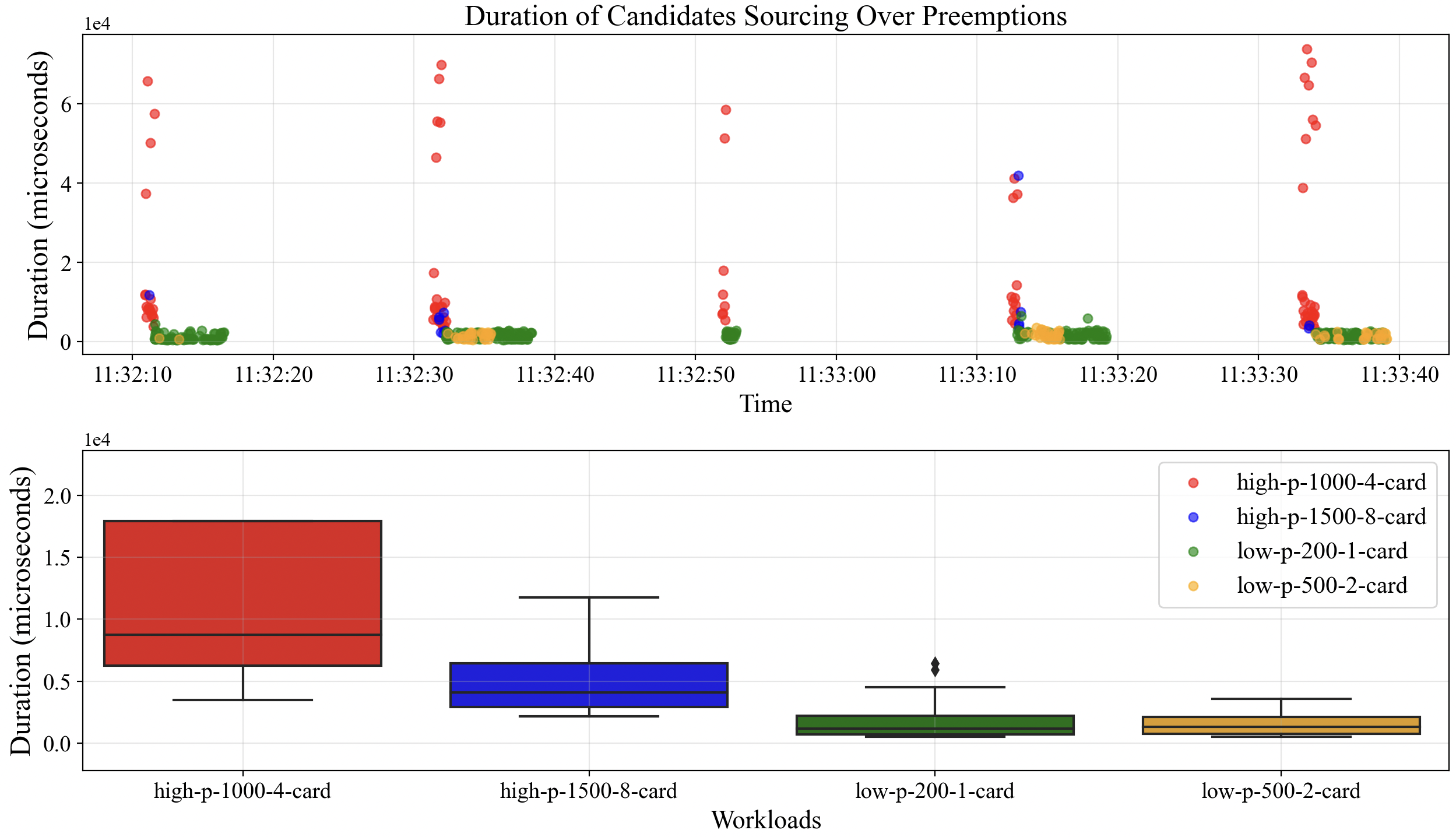}
    \caption{Overhead of candidate sourcing across various workloads}
    \label{fig:inter_workload_duration}
\end{figure}

First, we examine the overhead differences in candidate sourcing across various workloads. Figure~\ref{fig:inter_workload_duration} displays the time cost of candidate sourcing over five preemptions for each of the four workload types from Table~\ref{tab:workload-for-preemption}. Notably, Workload B incurs a significantly higher overhead in identifying suitable candidates compared to the other workloads. This is intuitive, as instances of Workload B require 4 GPUs, compelling the scheduler to evaluate numerous combinations of potential victims, especially those with only 1–2 GPU devices. Consequently, considerable time is spent assessing various combinations.

Conversely, Workload C, which requires only 2 GPUs per instance, has a notably lower time cost, validating this conclusion. Interestingly, Workload A, requiring 8 GPUs, incurs lower overhead than Workload B. This occurs because candidate sourcing for an 8-GPU preemptor terminates earlier due to quick failures in evaluating smaller victim combinations, resulting in fewer evaluated candidates than for Workload B’s 4-GPU requirement.

\begin{table}[!t]
\setlength{\tabcolsep}{12pt}
    \captionsetup{skip=5pt} 
    \centering
    \caption{Latency in microseconds on candidates sourcing}
    \renewcommand{\arraystretch}{1.2} 
    \begin{tabular}{lllr}
        \toprule
        \textbf{Workload} & \textbf{Method} & \textbf{P50} & \textbf{P90} \\
        \midrule
        \multirow{3}{*}{\textbf{high-p-1000-4-card}} 
        & Gödel            &  6421.0      &  62308.6  \\
        & FlexTopo         & 268011.0     & 416780.6  \\
        & FlexTopo-IMP      & 180019.5     & 275575.8  \\
        & \textbf{IMP Opt.} &  \textbf{32.8 $\%$}   &  \textbf{33.9 $\%$} \\
        \midrule
        \multirow{3}{*}{\textbf{low-p-500-2-card}} 
        & Gödel        &  1801.0  &  2491.5  \\
        & FlexTopo     &  2282.0  & 20624.6  \\
        & FlexTopo-IMP &  2115.0  &  4853.1  \\
        & \textbf{IMP Opt.} &  \textbf{7.3 $\%$}    & \textbf{76.5 $\%$} \\
        \bottomrule
    \end{tabular}
    \label{tab:latency-in-microseconds}
\end{table}

\begin{figure}[!h]
    \centering
    \includegraphics[width=0.88\linewidth]{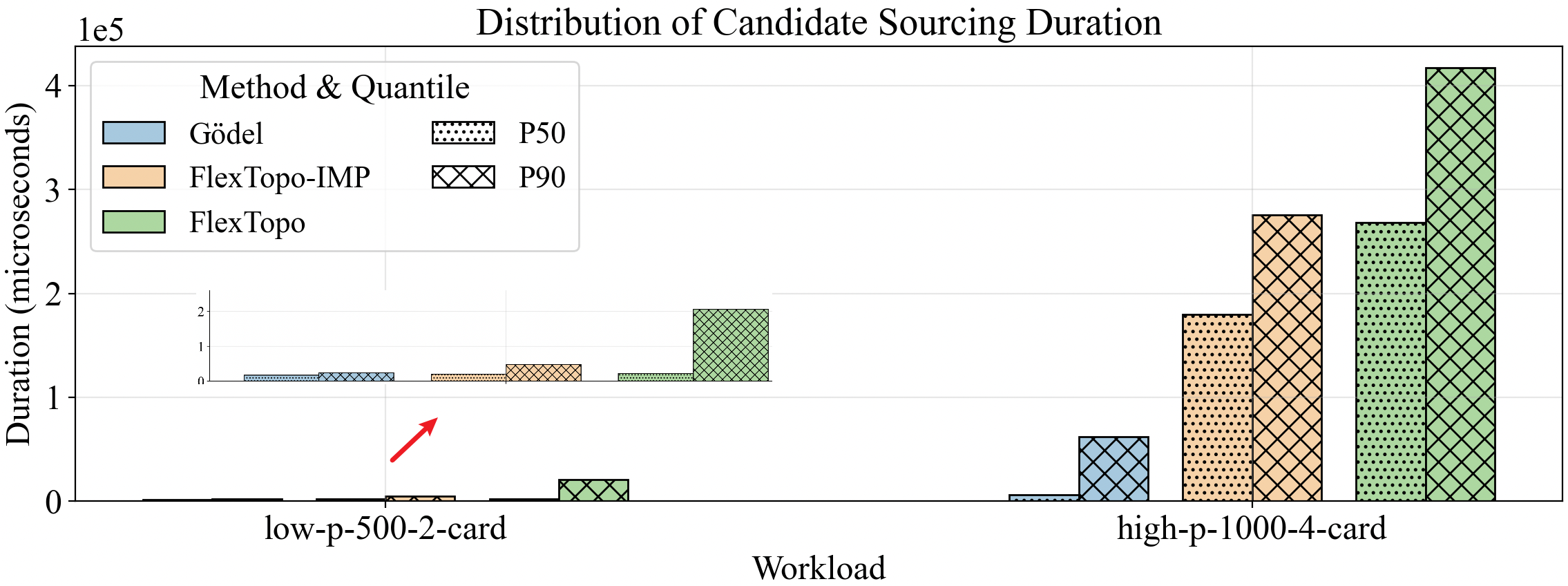}
    \caption{Distribution of candidate sourcing overhead in microseconds}
    \label{fig:distribution_of_three_methods}
\end{figure}

Second, we examine the overall time cost of topology-aware preemption across 5000 simulations. Table~\ref{tab:latency-in-microseconds} and Figure~\ref{fig:distribution_of_three_methods} present the results for Workloads B and C, comparing the candidate sourcing overhead and reporting the $P50$ and $P90$ latency in microseconds, respectively. As shown in Figure~\ref{fig:distribution_of_three_methods}, Gödel’s standard preemption achieves the best performance in all cases. This is because, in Gödel’s standard preemption policy, the scheduler directly selects the first feasible set of victims for each node, resulting in an almost constant, minimal time cost per node.

In contrast, FlexTopo-based preemption without IMP optimization introduces a substantial additional overhead, as expected, since the scheduler must exhaustively evaluate candidate solutions to find the optimal one. However, with the IMP algorithm, this overhead is significantly reduced by $7.3\%$ to $76.5\%$, closely approximating the time cost of Gödel’s standard preemption for Workload C. This efficiency gain results from the lower number of victims (2 in this case) required by each instance of Workload C. The IMP algorithm’s early-stopping mechanism effectively limits candidate exploration, significantly reducing the search cost.

\section{Related Work}
\label{sec:related-work}
\textbf{Workload Co-location and GPU-Sharing.}
Server-level workload co-location and device-level GPU sharing have shown significant potential in improving GPU utilization \cite{yeung2020towards, yeung2021horus, weng2023beware}. For instance, Horus \cite{yeung2021horus} introduces an interference-aware, prediction-based resource manager that refines workload co-location strategies by forecasting GPU utilization based on computation graph of deep learning model. While such approaches improve resource usage during workload co-location, they are not designed for preemptive scheduling, rendering them unsuitable for addressing preemption-based auto-scaling challenges under saturation allocation as discussed in Section~\ref{subsec:overview}. Device-level GPU sharing studies \cite{weng2023beware, zhao2020hived, ravi2011supporting, yu2020fine, goswami2016gpushare, nvidia2024gpusharing} concentrate on optimal workload placement on a single GPU to maintain SLAs while minimizing resource fragmentation. These works complement our approach by focusing on intra-device resource allocation.

\textbf{Topology-aware Scheduling.}
Amaral et al. \cite{amaral2017topology} present a topology-aware workload placement strategy for scheduling deep learning jobs on multi-GPU systems. The authors propose a graph-based topology representation to describe the communication patterns of jobs and the GPU distribution of hardware. This topology graph is then leveraged for informed scheduling decisions. However, the comprehensive topology representation is highly specific to certain GPU servers, such as the IBM Power8 system \cite{sinharoy2015advanced} equipped with 4 NVIDIA Tesla P100 cards \cite{cui40performance}. Additionally, the study does not explore adaptation and application in preemptive scheduling, where careful candidate and victim selection is crucial. 
Li et al. \cite{li2023topology} propose a Microservice-Oriented Topology-Aware Scheduling Framework (MOTAS), which utilizes microservice and cluster topologies to optimize network overhead in microservice applications using a heuristic graph mapping algorithm. Compared to MOTAS, our approach focuses on minimizing communication costs within the same server and can thus be used alongside MOTAS, which primarily addresses network overhead optimization. 
Gödel \cite{xiang2023godel} also supports topology-aware scheduling in both the normal scheduling cycle and preemption. It leverages the Custom Node Resource (CNR) CRD from the Katalyst project \cite{katalyst2024core}, open-sourced by ByteDance, to represent the topology of each server in the cluster. However, Gödel's NUMA-aware preemption mainly supports \textit{Exclusive Dedicated Cores}, \textit{Non-Exclusive Dedicated Cores}, and \textit{Shared Cores} NUMA allocation, rather than offering flexible topology affinity. These three NUMA policies need to be tailored to the specific server hardware and workload requirements individually, posing challenges for application in large-scale clusters. Another line of work \cite{lee2010topology, ryu2020towards, han2022microsecond} explores the use of topology information in specialized application scenarios such as autonomous driving and virtual reality. In contrast, our research focuses on LLM serving within co-located cloud-native workloads.

\textbf{LLM Inference Optimization.}
Many studies have aimed to enhance LLM inference efficiency. One focus is on memory optimization, using techniques like KV cache management and reuse \cite{zheng2023efficiently, kwon2023efficient}. Another line of research targets latency reduction through scheduling improvements \cite{yu2022orca, wu2023fast}, early exits \cite{zhou2020bert}, and kernel optimization \cite{dao2022flashattention}. Further approaches include sparsification \cite{frantar2023sparsegpt} and quantization \cite{lin2024awq}. Unlike these methods that optimize inference engine itself, our work keeps the engine setup fixed and emphasizes on victim selection for optimal instance placement during preemptive auto-scaling. These line of work is complementary with ours.

\section{Conclusion}
\label{sec:conclusion}
In this study, we presented a topology-aware preemptive scheduling approach tailored to the demands of LLM workloads in co-located clusters. Traditional schedulers often lack topology awareness, leading to suboptimal preemption that fails to meet the high-performance needs of latency-sensitive, high-priority services. Our solution, FlexTopo-based IMP, provides a unified, flexible topology representation to guide preemption decisions, ensuring that resources freed by preempted tasks match the topological preferences of critical workloads, thereby improving efficiency and reducing resource fragmentation.

Evaluations on simulated and near-production clusters demonstrate FlexTopo’s effectiveness, achieving a $100\%$ topology affinity hit rate versus $45\%$ with standard methods, reducing scheduling failures and improving scheduled performance by $55\%$. This method not only enhances resource elasticity and efficiency but also offers a scalable framework for topology-aware scheduling in dynamic, large-scale environments.

\bibliographystyle{unsrt}  
\bibliography{references}  

\begin{thebibliography}{10}

\bibitem{hadi2023survey}
Muhammad~Usman Hadi, Rizwan Qureshi, Abbas Shah, Muhammad Irfan, Anas Zafar,
  Muhammad~Bilal Shaikh, Naveed Akhtar, Jia Wu, Seyedali Mirjalili, et~al.
\newblock A survey on large language models: Applications, challenges,
  limitations, and practical usage.
\newblock {\em Authorea Preprints}, 2023.

\bibitem{duan2024efficient}
Jiangfei Duan, Shuo Zhang, Zerui Wang, Lijuan Jiang, Wenwen Qu, Qinghao Hu,
  Guoteng Wang, Qizhen Weng, Hang Yan, Xingcheng Zhang, et~al.
\newblock Efficient training of large language models on distributed
  infrastructures: A survey.
\newblock {\em arXiv preprint arXiv:2407.20018}, 2024.

\bibitem{wang2024towards}
Yuxin Wang, Yuhan Chen, Zeyu Li, Zhenheng Tang, Rui Guo, Xin Wang, Qiang Wang,
  Amelie~Chi Zhou, and Xiaowen Chu.
\newblock Towards efficient and reliable llm serving: A real-world workload
  study.
\newblock {\em arXiv preprint arXiv:2401.17644}, 2024.

\bibitem{xiang2023godel}
Wu~Xiang, Yakun Li, Yuquan Ren, Fan Jiang, Chaohui Xin, Varun Gupta, Chao
  Xiang, Xinyi Song, Meng Liu, Bing Li, et~al.
\newblock G{\"o}del: Unified large-scale resource management and scheduling at
  bytedance.
\newblock In {\em Proceedings of the 2023 ACM Symposium on Cloud Computing},
  pages 308--323, 2023.

\bibitem{amaral2017topology}
Marcelo Amaral, Jord{\`a} Polo, David Carrera, Seetharami Seelam, and
  Malgorzata Steinder.
\newblock Topology-aware gpu scheduling for learning workloads in cloud
  environments.
\newblock In {\em Proceedings of the International Conference for High
  Performance Computing, Networking, Storage and Analysis}, pages 1--12, 2017.

\bibitem{lameter2013numa}
Christoph Lameter.
\newblock Numa (non-uniform memory access): An overview: Numa becomes more
  common because memory controllers get close to execution units on
  microprocessors.
\newblock {\em Queue}, 11(7):40--51, 2013.

\bibitem{griggs2024m}
Tyler Griggs, Xiaoxuan Liu, Jiaxiang Yu, Doyoung Kim, Wei-Lin Chiang, Alvin
  Cheung, and Ion Stoica.
\newblock M$\backslash$'elange: Cost efficient large language model serving by
  exploiting gpu heterogeneity.
\newblock {\em arXiv preprint arXiv:2404.14527}, 2024.

\bibitem{nvidia2021fastertransformer}
NVIDIA.
\newblock Fastertransformer.
\newblock \url{https://github.com/NVIDIA/FasterTransformer}, 2021.
\newblock Commit: df4a753, Accessed on: 2023-11-25.

\bibitem{kwon2023efficient}
Woosuk Kwon, Zhuohan Li, Siyuan Zhuang, Ying Sheng, Lianmin Zheng, Cody~Hao Yu,
  Joseph Gonzalez, Hao Zhang, and Ion Stoica.
\newblock Efficient memory management for large language model serving with
  pagedattention.
\newblock In {\em Proceedings of the 29th Symposium on Operating Systems
  Principles}, pages 611--626, 2023.

\bibitem{flexgen2023}
FMInference.
\newblock Flexgen.
\newblock \url{https://github.com/FMInference/FlexGen}, 2023.
\newblock Commit: d34f7b4, Accessed on: 2023-11-25.

\bibitem{huggingface2023textgen}
Huggingface.
\newblock Huggingface text generation inference.
\newblock \url{https://github.com/huggingface/text-generation-inference}, 2023.
\newblock Commit: 3c02262, Accessed on: 2023-11-25.

\bibitem{deepspeed2022inference}
Microsoft.
\newblock Deepspeed inference.
\newblock \url{https://github.com/microsoft/DeepSpeed}, 2022.
\newblock Commit: 2afa1c7, Accessed on: 2023-11-25.

\bibitem{nvidia2023tensorrtllm}
NVIDIA.
\newblock Tensorrt-llm.
\newblock \url{https://github.com/NVIDIA/TensorRT-LLM}, 2023.
\newblock Commit: 6837c81, Accessed on: 2023-11-25.

\bibitem{miao2023towards}
Xupeng Miao, Gabriele Oliaro, Zhihao Zhang, Xinhao Cheng, Hongyi Jin, Tianqi
  Chen, and Zhihao Jia.
\newblock Towards efficient generative large language model serving: A survey
  from algorithms to systems.
\newblock {\em arXiv preprint arXiv:2312.15234}, 2023.

\bibitem{Kubernetes2020topologymanager}
Kubernetes.io.
\newblock Kubernetes topology manager moves to beta.
\newblock
  \url{https://kubernetes.io/blog/2020/04/01/kubernetes-1-18-feature-topoloy-manager-beta/},
  2023.
\newblock Published on: 2020-04-01.

\bibitem{preemption2024k8s}
Kubernetes.io.
\newblock Pod priority and preemption.
\newblock
  \url{https://kubernetes.io/docs/concepts/scheduling-eviction/pod-priority-preemption},
  2024.
\newblock Accessed on: 2024-11-04.

\bibitem{kubewharfGodel-scheduler}
Kubewharf.
\newblock Godel scheduler: a unified scheduler for online and offline tasks.
\newblock \url{https://github.com/kubewharf/godel-scheduler}, 2023.

\bibitem{daemonset}
Kubernetes.io.
\newblock Daemonset.
\newblock
  \url{https://kubernetes.io/docs/concepts/workloads/controllers/daemonset/},
  2024.
\newblock Accessed on: 2024-11-14.

\bibitem{kwok2024k8s}
Kwok.sigs.k8s.io.
\newblock Kubernetes without kubelet.
\newblock \url{https://kwok.sigs.k8s.io/}, 2024.
\newblock Accessed on: 2024-11-13.

\bibitem{topoerrork8s}
Kubernetes.io.
\newblock Control topology management policies on a node.
\newblock
  \url{https://kubernetes.io/docs/tasks/administer-cluster/topology-manager/policy-restricted},
  2024.
\newblock Accessed on: 2024-11-14.

\bibitem{yeung2020towards}
Gingfung Yeung, Damian Borowiec, Adrian Friday, Richard Harper, and Peter
  Garraghan.
\newblock Towards $\{$GPU$\}$ utilization prediction for cloud deep learning.
\newblock In {\em 12th USENIX Workshop on Hot Topics in Cloud Computing
  (HotCloud 20)}, 2020.

\bibitem{yeung2021horus}
Gingfung Yeung, Damian Borowiec, Renyu Yang, Adrian Friday, Richard Harper, and
  Peter Garraghan.
\newblock Horus: Interference-aware and prediction-based scheduling in deep
  learning systems.
\newblock {\em IEEE Transactions on Parallel and Distributed Systems},
  33(1):88--100, 2021.

\bibitem{weng2023beware}
Qizhen Weng, Lingyun Yang, Yinghao Yu, Wei Wang, Xiaochuan Tang, Guodong Yang,
  and Liping Zhang.
\newblock Beware of fragmentation: Scheduling $\{$GPU-Sharing$\}$ workloads
  with fragmentation gradient descent.
\newblock In {\em 2023 USENIX Annual Technical Conference (USENIX ATC 23)},
  pages 995--1008, 2023.

\bibitem{zhao2020hived}
Hanyu Zhao, Zhenhua Han, Zhi Yang, Quanlu Zhang, Fan Yang, Lidong Zhou, Mao
  Yang, Francis~CM Lau, Yuqi Wang, Yifan Xiong, et~al.
\newblock $\{$HiveD$\}$: Sharing a $\{$GPU$\}$ cluster for deep learning with
  guarantees.
\newblock In {\em 14th USENIX symposium on operating systems design and
  implementation (OSDI 20)}, pages 515--532, 2020.

\bibitem{ravi2011supporting}
Vignesh~T Ravi, Michela Becchi, Gagan Agrawal, and Srimat Chakradhar.
\newblock Supporting gpu sharing in cloud environments with a transparent
  runtime consolidation framework.
\newblock In {\em Proceedings of the 20th international symposium on High
  performance distributed computing}, pages 217--228, 2011.

\bibitem{yu2020fine}
Peifeng Yu and Mosharaf Chowdhury.
\newblock Fine-grained gpu sharing primitives for deep learning applications.
\newblock {\em Proceedings of Machine Learning and Systems}, 2:98--111, 2020.

\bibitem{goswami2016gpushare}
Anshuman Goswami, Jeffrey Young, Karsten Schwan, Naila Farooqui, Ada
  Gavrilovska, Matthew Wolf, and Greg Eisenhauer.
\newblock Gpushare: Fair-sharing middleware for gpu clouds.
\newblock In {\em 2016 IEEE International Parallel and Distributed Processing
  Symposium Workshops (IPDPSW)}, pages 1769--1776. IEEE, 2016.

\bibitem{nvidia2024gpusharing}
NVIDIA.
\newblock Nvidia cloud native technologies: Gpu sharing.
\newblock
  \url{https://docs.nvidia.com/datacenter/cloud-native/gpu-operator/latest/gpu-sharing.html},
  2024.
\newblock Accessed on: 2024-11-04.

\bibitem{sinharoy2015advanced}
Balaram Sinharoy, Randy Swanberg, Naresh Nayar, B~Mealey, Jeff Stuecheli, Berni
  Schiefer, Jens Leenstra, Joefon Jann, Philipp Oehler, David Levitan, et~al.
\newblock Advanced features in ibm power8 systems.
\newblock {\em IBM Journal of Research and Development}, 59(1):1--1, 2015.

\bibitem{cui40performance}
Xuewen Cui, Thomas~RW Scogland, Bronis~R de~Supinski, and Wu-chun Feng.
\newblock Performance evaluation of the nvidia tesla p100: Our directive-based
  partitioning and pipelining vs. nvidia’s unified memory.
\newblock {\em Matrix}, 40:50, 2017.

\bibitem{li2023topology}
Xin Li, Junsong Zhou, Xin Wei, Dawei Li, Zhuzhong Qian, Jie Wu, Xiaolin Qin,
  and Sanglu Lu.
\newblock Topology-aware scheduling framework for microservice applications in
  cloud.
\newblock {\em IEEE Transactions on Parallel and Distributed Systems},
  34(5):1635--1649, 2023.

\bibitem{katalyst2024core}
Kubewharf.
\newblock Katalyst core.
\newblock \url{https://github.com/kubewharf/katalyst-core}, 2023.
\newblock Accessed on: 2024-11-04.

\bibitem{lee2010topology}
Gunho Lee, Niraj Tolia, Parthasarathy Ranganathan, and Randy~H Katz.
\newblock Topology-aware resource allocation for data-intensive workloads.
\newblock In {\em Proceedings of the first ACM asia-pacific workshop on
  Workshop on systems}, pages 1--6, 2010.

\bibitem{ryu2020towards}
Bon Ryu, Aijun An, Zana Rashidi, Junfeng Liu, and Yonggang Hu.
\newblock Towards topology aware pre-emptive job scheduling with deep
  reinforcement learning.
\newblock In {\em Proceedings of the 30th Annual International Conference on
  Computer Science and Software Engineering}, pages 83--92, 2020.

\bibitem{han2022microsecond}
Mingcong Han, Hanze Zhang, Rong Chen, and Haibo Chen.
\newblock Microsecond-scale preemption for concurrent
  $\{$GPU-accelerated$\}$$\{$DNN$\}$ inferences.
\newblock In {\em 16th USENIX Symposium on Operating Systems Design and
  Implementation (OSDI 22)}, pages 539--558, 2022.

\bibitem{zheng2023efficiently}
Lianmin Zheng, Liangsheng Yin, Zhiqiang Xie, Jeff Huang, Chuyue Sun, Cody
  Hao~Yu, Shiyi Cao, Christos Kozyrakis, Ion Stoica, Joseph~E Gonzalez, et~al.
\newblock Efficiently programming large language models using sglang.
\newblock {\em arXiv e-prints}, pages arXiv--2312, 2023.

\bibitem{yu2022orca}
Gyeong-In Yu, Joo~Seong Jeong, Geon-Woo Kim, Soojeong Kim, and Byung-Gon Chun.
\newblock Orca: A distributed serving system for $\{$Transformer-Based$\}$
  generative models.
\newblock In {\em 16th USENIX Symposium on Operating Systems Design and
  Implementation (OSDI 22)}, pages 521--538, 2022.

\bibitem{wu2023fast}
Bingyang Wu, Yinmin Zhong, Zili Zhang, Gang Huang, Xuanzhe Liu, and Xin Jin.
\newblock Fast distributed inference serving for large language models.
\newblock {\em arXiv preprint arXiv:2305.05920}, 2023.

\bibitem{zhou2020bert}
Wangchunshu Zhou, Canwen Xu, Tao Ge, Julian McAuley, Ke~Xu, and Furu Wei.
\newblock Bert loses patience: Fast and robust inference with early exit.
\newblock {\em Advances in Neural Information Processing Systems},
  33:18330--18341, 2020.

\bibitem{dao2022flashattention}
Tri Dao, Dan Fu, Stefano Ermon, Atri Rudra, and Christopher R{\'e}.
\newblock Flashattention: Fast and memory-efficient exact attention with
  io-awareness.
\newblock {\em Advances in Neural Information Processing Systems},
  35:16344--16359, 2022.

\bibitem{frantar2023sparsegpt}
Elias Frantar and Dan Alistarh.
\newblock Sparsegpt: Massive language models can be accurately pruned in
  one-shot.
\newblock In {\em International Conference on Machine Learning}, pages
  10323--10337. PMLR, 2023.

\bibitem{lin2024awq}
Ji~Lin, Jiaming Tang, Haotian Tang, Shang Yang, Wei-Ming Chen, Wei-Chen Wang,
  Guangxuan Xiao, Xingyu Dang, Chuang Gan, and Song Han.
\newblock Awq: Activation-aware weight quantization for on-device llm
  compression and acceleration.
\newblock {\em Proceedings of Machine Learning and Systems}, 6:87--100, 2024.

\end{thebibliography}
\end{document}